\documentclass{llncs}

\usepackage{graphicx}
\usepackage{subcaption}

\usepackage[english]{babel}
\usepackage{amsmath}
\usepackage{amsfonts}
\usepackage{amssymb}
\usepackage{paralist}
\usepackage{bussproofs}
\usepackage{stmaryrd}
\usepackage{color}
\usepackage{mathpartir}
\usepackage{algorithm}
\usepackage[noend]{algpseudocode} 
\usepackage[
   a4paper,
   pdftex,
   pdftitle={The Imandra Theorem Prover (system description)},
   pdfauthor={Grant Passmore, Simon Cruanes},
   pdfkeywords={},
   pdfborder={0 0 0},
   draft=false,
   bookmarksnumbered,
   bookmarks,
   bookmarksdepth=2,
   bookmarksopenlevel=2,
   bookmarksopen]{hyperref}
\usepackage{enumitem}  
\usepackage[final]{listings}
\usepackage{extarrows} 
\usepackage{booktabs} 
\usepackage{cite} 

\setitemize{noitemsep,topsep=0pt,parsep=0pt,partopsep=0pt}

\newif\ifInReport
\InReportfalse

\newcommand\todoEnabled{1}
\newcommand{\todo}[2][red]{
    \PackageWarning{TODO}{#2}
    \ifx\todoEnabled\undefined{}\else{
    \marginpar{
      \fbox{
        \begin{minipage}{.2\textwidth}
            \begin{flushleft}
                \footnotesize\textcolor{#1}{TODO$_{(\the\inputlineno)}$~#2}
            \end{flushleft}
        \end{minipage}
      }
    }
    }\fi
}

\newcommand\eqdef{\ensuremath{\mathrel{\overset{\text{\tiny def}}{=}} }}
\newcommand{\bigAnd}{\ensuremath{\bigwedge}}

\newcommand{\limply}{\ensuremath{\mathrel{\Rightarrow} }}

\newcommand\subterm{\ensuremath{\triangleleft}}

\newcommand\true{\ensuremath\top}
\newcommand\false{\ensuremath\bot}

\newcommand\ite[3]{\ensuremath{\textsf{if}~#1~#2~#3}}

\newcommand\factSym{\ensuremath{\textsf{fact} }}
\newcommand\fact[1]{\ensuremath{\factSym(#1)}}
\newcommand\blockLit[1]{\ensuremath{\textsf{b}[#1]} }

\begin{document}

\ifInReport
\else
\title{The Imandra Automated Reasoning System (system description)}
\fi

\titlerunning{Imandra}
\authorrunning{Grant Olney Passmore, Simon Cruanes, Team Imandra}
\author{Grant Olney Passmore, Simon Cruanes, Denis Ignatovich, Dave Aitken, Matt Bray, Elijah Kagan, Kostya Kanishev, Ewen Maclean, Nicola Mometto}
\institute{ Imandra Inc., USA }

\maketitle

\begin{abstract}
We describe Imandra, a modern computational logic theorem prover designed to
bridge the gap between decision procedures such as SMT, semi-automatic inductive
provers of the Boyer-Moore family like ACL2, and interactive proof assistants
for typed higher-order logics.
Imandra's logic is computational, based on a pure subset of OCaml in which all
functions are terminating, with restrictions on types and higher-order functions
that allow conjectures to be translated into multi-sorted first-order logic with
theories, including arithmetic and datatypes.
Imandra has novel features supporting large-scale industrial applications,
including a seamless integration of bounded and unbounded verification,
first-class computable counterexamples, efficiently executable models and a
cloud-native architecture supporting live multiuser collaboration.
The core reasoning mechanisms of Imandra are
\begin{inparaenum}[(i)]
\item a semi-complete procedure for finding models of formulas in the logic
  mentioned above, centered around the lazy expansion of recursive functions,
\item an inductive waterfall and simplifier which ``lifts'' many Boyer-Moore
  ideas to our typed higher-order setting.
\end{inparaenum}
These mechanisms are tightly integrated and subject to many forms of user control.
\end{abstract}

\section{Introduction}

Imandra is a modern computational logic theorem prover built around a pure,
higher-order subset of OCaml.
Mathematical models and conjectures are written as executable OCaml programs,
and Imandra may be used to reason about them, combining models, proofs and
counterexamples in a unified computational environment.
Imandra is designed to bridge the gap between decision procedures such as
SMT~\cite{z3_prover}, semi-automatic inductive provers of the Boyer-Moore family
like ACL2~\cite{boyer-moore-acl,kaufmann1996acl2}, and interactive proof
assistants for typed higher-order
logics~\cite{gordon-melham-hol,nipkow-paulson-wenzel:2002,harrison-hollight,cade92-pvs}.
Our goal is to build a friendly, easy to use system by leveraging strong
automation in proof search that can also robustly provide counterexamples
for false conjectures.
Imandra has novel features supporting large-scale industrial applications,
including a seamless integration of bounded and unbounded verification,
first-class computable counterexamples, efficiently executable models and a
cloud-native architecture supporting live multiuser collaboration.
Imandra is already in use by major companies in the financial sector, including
Goldman Sachs, Itiviti and OneChronos~\cite{passmore2017formal}.

An online version may be found at \url{https://try.imandra.ai}.
\begin{figure}
  \centering
  \vspace{-20pt}
  \includegraphics[width=1\textwidth]{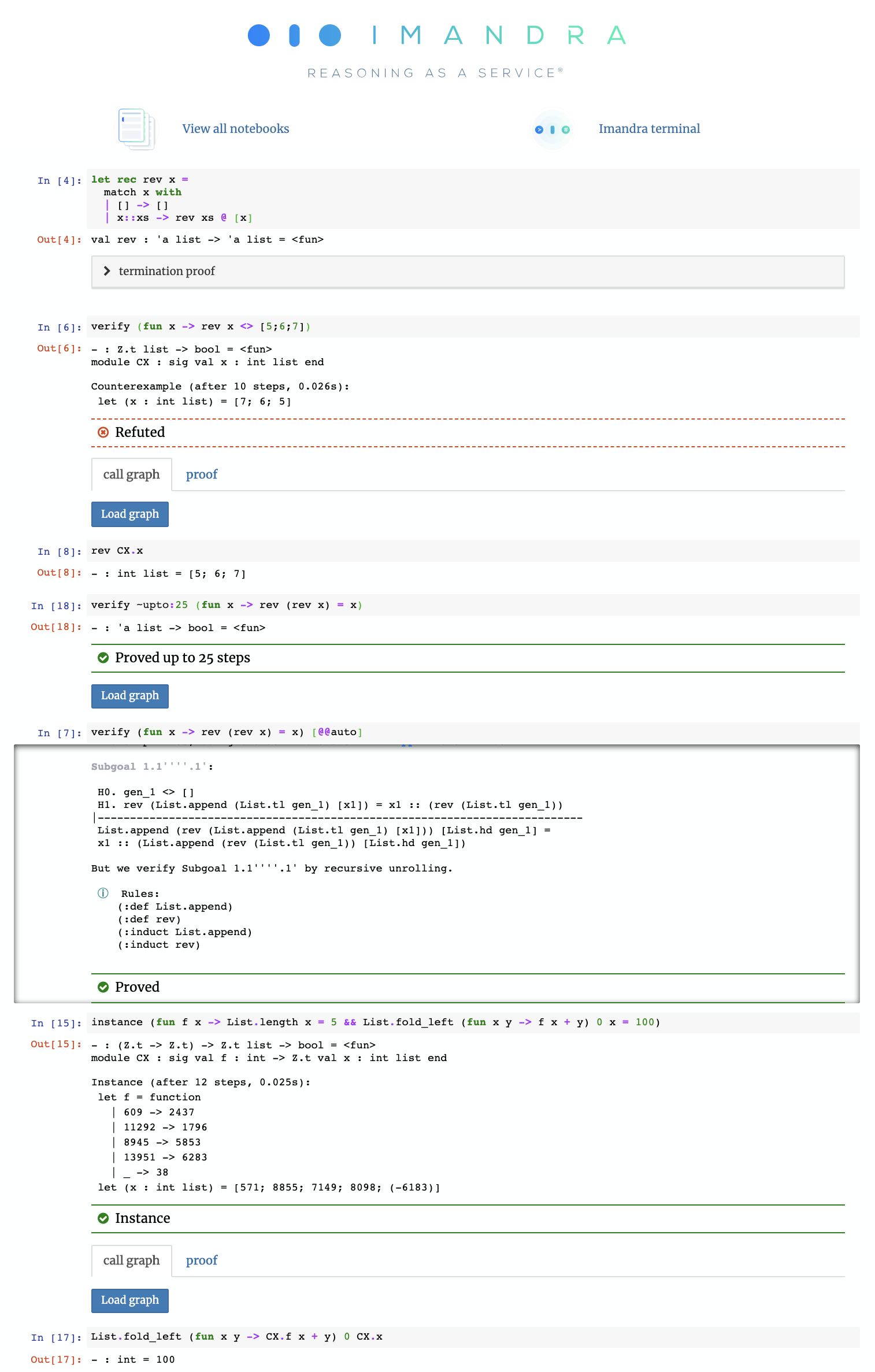}
    \caption{An example Imandra session illustrating recursive definitions,
      computable counterexamples (\lstinline{CX}), bounded verification
      (\lstinline{verify upto}), unbounded verification with automated induction
      (\lstinline{@@auto}), and higher-order instance synthesis.}
    \label{fig:screen:jymandra}
\end{figure}


\section{Logic}\label{sec:logic}

Imandra's logic is built on a mechanized formal semantics for a pure,
higher-order subset of OCaml.
Foundationally, the subset of OCaml Imandra supports (called the `Imandra
Modelling Language') corresponds to a (specializable) computational
fragment of HOL equivalent to multi-sorted first-order logic with induction up to
$\epsilon_0$ extended with theories of datatypes, integer and real arithmetic.
Theorems are implicitly universally quantified and expressed as Boolean-valued
functions.
Proving a theorem establishes that the corresponding function always evaluates
to \lstinline{true}.
As in PRA (Primitive Recursive Arithmetic) and Boyer-Moore logics,
existential goals are expressed with explicit
computable Skolem functions~\cite{goodstein-rnt, skolem-pra, boyer-moore-acl}.

\subsection{Definitional Principle}\label{sec:termination}

Users work with Imandra by incrementally extending its logical world through
definitions of types, functions, modules and theorems.
Each extension is governed by a definitional principle designed to maintain the
consistency of Imandra's current logical theory through a discipline of
\emph{conservative extensions}.
Types must be proved well-founded.
Functions must be proved terminating.
These termination proofs play a dual role: Their structure is mined in order to
instruct Imandra how to construct induction principles tailored to the recursive
function being admitted when it later appears in conjectures.

Imandra's definitional principle is built upon the ordinals up to $\epsilon_0$.
Ordinals are encoded as a datatype (\lstinline{Ordinal.t}) in Imandra using a
variant of Cantor normal form, and the well-foundedness of
\lstinline{Ordinal.(<<)} --- the strict less-than relation on \lstinline{Ordinal.t}
values --- is an axiom of Imandra's logic.

To prove a function $f$ terminating, an ordinal-valued {\em measure} is
required.
Measures can often be inferred (e.g., for structural recursions) and may 
be specified by the user.
To establish termination, all recursive calls of $f$ are collected together with
their guards, and their arguments must be proved to conditionally map to
strictly smaller ordinals via the measure.
Imandra provides a shorthand annotation for specifying lexicographic orders
(\lstinline{@@adm}), and explicit measure functions may be given using the
\lstinline{@@measure} annotation.

\begin{example}[Ackermann]
We can define the Ackermann function and prove it terminating with the
attribute \lstinline{[@@adm m,n]} which maps \lstinline{ack m n} to the ordinal
$m\cdot \omega + n$.
Alternatively, we could use \lstinline{[@@measure Ordinal.(pair (of_int m) (of_int n))]}
to give an explicit measure via helper functions in Imandra's
\lstinline{Ordinal} module.
\begin{lstlisting}
let rec ack m n =
  if m <= 0 then n + 1 else if n <= 0 then ack (m-1) 1 else ack (m-1) (ack m (n-1))
[@@adm m,n]
\end{lstlisting}
%
\end{example}
\begin{example}
Here we have a naive version of the classic
{\em left-pad}
function~\cite{leftpad}, where termination depends on both arguments in a non-lexicographic manner:
\begin{lstlisting}
let rec left_pad c n xs =
  if List.length xs >= n then xs else left_pad c n (c :: xs)
[@@measure Ordinal.of_int (n - List.length xs)]
\end{lstlisting}
\end{example}

\subsection{Lifting, Specialization and Monomorphization}

Imandra definitions may be polymorphic and higher-order.
However, once Imandra is tasked with determining the truth value of a
conjecture, the goal and its transitive dependencies are transformed into a
family of ground, monomorphic first-order (recursive) definitions.
These transformations include lambda lifting, specialization and monomorphization.
Imandra's supported fragment of OCaml is designed so that all admitted definitions
may be transformed in this way. 

\begin{example}
To prove the following higher-order theorem
\begin{lstlisting}
theorem same_len l =
  List.length (List.map (fun x -> x+1) l) = List.length l
\end{lstlisting}
\noindent we obtain a set of lower level definitions, where the anonymous
function was lifted, the type \lstinline{list} was monomorphised, and
\lstinline{map} and \lstinline{length} were specialised:

\begin{lstlisting}
type int_list = Nil_int | Cons_int of int * int_list
let rec length_int = function
  | Nil_int -> 0
  | Cons_int (_, tl) -> 1 + length_int tl
let map_lambda0 x = x+1
let rec map1 = function
  | Nil_int -> Nil_int
  | Cons_int (x, tl) -> Cons_int (map_lambda0 x, map1 tl)
theorem same_len (l:int_list) : bool =
  length_int (map1 l) = length_int l
\end{lstlisting}
\end{example}



\section{Unrolling of Recursive Functions}\label{sec:recfun}

A major feature of Imandra is its ability to automatically search for proofs and
counterexamples in a logic with recursive functions.
When a counterexample is found, it is reflected as a first-class value in
Imandra's runtime and can be directly computed with and run through the model
being analysed.
In fact, the statement \lstinline{verify (fun x -> ...)} does not try any
inductive proving unless requested; the default strategy is recursive function
{\em unrolling} for a fixed number of steps, a form of bounded symbolic model-checking.

Our core unrolling algorithm is similar in spirit to the work of Suter et
al.~\cite{suter2011satisfiability} but with crucial strategic differences.
In essence, Imandra uses the {\em assumption} mechanism of SMT to block all
Boolean assignments that involve the evaluation of a (currently) uninterpreted
ground instance of a recursive function.
A refinement loop, based on extraction of unsat-cores from this set of
assumptions, then expands (interprets) the function calls one by one until a
model is found, an empty unsat-core is obtained, or a maximal number of steps is
reached.

\begin{definition}[Function template]
  A function template for $f$ is a set of tuples $(g, \vec{t}, \vec{p})$
  such that the body of $f$ contains a call to $g(\vec{t})$ under
  the path $\vec{p}$.
\end{definition}

\begin{example}
  ~\\
  \vspace{-10pt}
  \begin{itemize}[itemsep=.5pt]
  \item $\fact{x}= \ite{x > 1}{(x * \fact{x-1})}{1}$
    has as template 
    $\{ (\factSym, \vec{(x-1)}, (x>1)) \}$

  \item $f(x) = 1 + \ite{g(0)}{h(g(x))}{h(42)}$ \\
    has as template \\
      $\{ (g, (0), \true), (h, \vec{(g(x))}, (g(0)=\true)),
          (g, \vec{(x)}, g(0)=\true), (h, \vec{(42)}, (g(0)=\false))
      \}$
  \end{itemize}

\end{example}

We use what we call {\em reachability literals} to prevent the SMT solver from
picking assignments that use function calls that are not expanded yet.
A reachability literal is a Boolean atom that doesn't appear in the original
problem, and that we associate to a given function call $f(\vec{t})$ regardless
of where it occurs.
This is to be contrasted with Suter et al.'s notion of 
\emph{control literals} associated with individual occurrences of function
calls within the expanded body of another function call.
We denote by $\blockLit{f(\vec{t})}$ the unique reachability literal for $f(\vec{t})$.

\begin{figure}
\begin{center}
\vspace{-10pt}
\begin{lstlisting}[language=python,numbers=left,xleftmargin=18pt,numberstyle=\small]
def calls_of_term(t: Term):
  return LTX $\{ b[f(\vec{u})] ~|~ f(\vec{u}) \subterm t \}$ LTX

def subcalls_of_call(LTX$f(\vec{t})$LTX: Term, expanded: Set[Term]):
  return LTX$\{ (b[g(\vec{u})], p) ~|~ (g, \vec{u}, \bigAnd p)
    \in \textsf{template}(f)[\vec{t}/\vec{x}] \land g(\vec{u})
    \not\in \textsf{expanded} \}$ LTX

def unroll(goal: Formula) -> SAT|UNSAT:
  q = calls_of_term(goal), expanded = LTX$\emptyset$LTX
  F = goal LTX$ ~\land \bigAnd_{ a \in q } a$LTX
  while True:
    is_sat, unsat_core = check_sat(F, assume=LTX$\{ \lnot a ~|~ a \in q \} $LTX)  LTX\label{code:check-sat}LTX
    if is_sat == SAT: return SAT
    else if is_sat == UNSAT:
      if unsat_core == LTX$\emptyset$LTX: return UNSAT
      LTX$ b[f(\vec{t})] $LTX = pick_from(unsat_core)  # next call to expand   LTX\label{code:pick-expand}LTX
      expanded = LTX$\{ f(\vec{t}) \} \cup \textsf{expanded}$LTX
      LTX$ \{ (a_i, p_i) \}_i $LTX = subcalls_of_call(LTX$f(\vec{t})$LTX, expanded)
      q = q LTX$ \cup ~\{ a_i \}_i \setminus b[f(\vec{t})] $LTX
      F = F LTX$ \land ~ b[f(\vec{t})]
        \land f(\vec{t}) = body_{f}[\vec{t}/\vec{x}]
        \land \bigAnd_i \left( b[f(\vec{t})] \land p_i \limply a_i \right)
        $LTX
\end{lstlisting}
\end{center}
\vspace{-20pt}
\caption{Unrolling algorithm}
\label{fig:unroll-loop}
\end{figure}
\vspace{-10pt}
The main search loop is presented in Figure~\ref{fig:unroll-loop},
where $body_f$ is the body of $f$ (i.e. $\smash{ f(\vec{x}) \eqdef body_f }$)
and $t \subterm u$ means $t$ is a proper subterm of $u$.
We start with $F$ initialized to the original goal, and the queue $q$ containing
function calls in the goal (computed by \lstinline{calls_of_term}).
Each iteration of the loop starts by checking validity under the assumption that all
reachability literals in $q$ are false (line~\ref{code:check-sat}).
If no model is found, we pick an unexpanded function call $f(\vec{t})$ from
the unsat core (line \ref{code:pick-expand}).
Selection must be \emph{fair}: all function calls must eventually be picked.

To expand $f(\vec{t})$, the corresponding reachability literal becomes true, we instantiate
the body of $f$ on $\vec{t}$, and use
\lstinline{subcalls_of_call} to compute the set of subcalls
along with their control path within $f(\vec{t})$ (using $f$'s template).
For each $\blockLit{g(\vec{u})}$ occurring under path $p$ inside $\smash{
  \textsf{body}_f(\vec{t}) }$, we need to block models that would make $p$ valid
until $g(\vec{u})$ gets expanded.
The assertions $\bigAnd_i \left( b[f(\vec{t})] \land p_i \limply a_i \right)$
delegate to SMT the work of tracking which paths are forbidden.
This way, expanding one function call might lead to many paths becoming
``unlocked'' at once.

\section{Induction}\label{sec:induction}

%
Imandra has extensive support for automated induction built principally around
Imandra's \emph{inductive waterfall}\footnote{More details about Imandra's
  waterfall and rule classes may be found in our online documentation at
  \url{https://docs.imandra.ai}.}.
This combines techniques such as symbolic execution, lemma-based
conditional rewriting, forward-chaining, generalization and the automatic
synthesis of goal-specific induction principles.
%
Induction principle synthesis depends upon data computed about a function's
termination obtained when it was admitted via our definitional principle.
Imandra's waterfall is deeply inspired by the pioneering work of
Boyer-Moore~\cite{kaufmann1996acl2,boyer-moore-acl}, and is in many ways a
``lifting'' of the Boyer-Moore waterfall to our typed, higher-order
setting.

\begin{figure}
  \centering
  \vspace{-10pt}
  \includegraphics[width=1\textwidth]{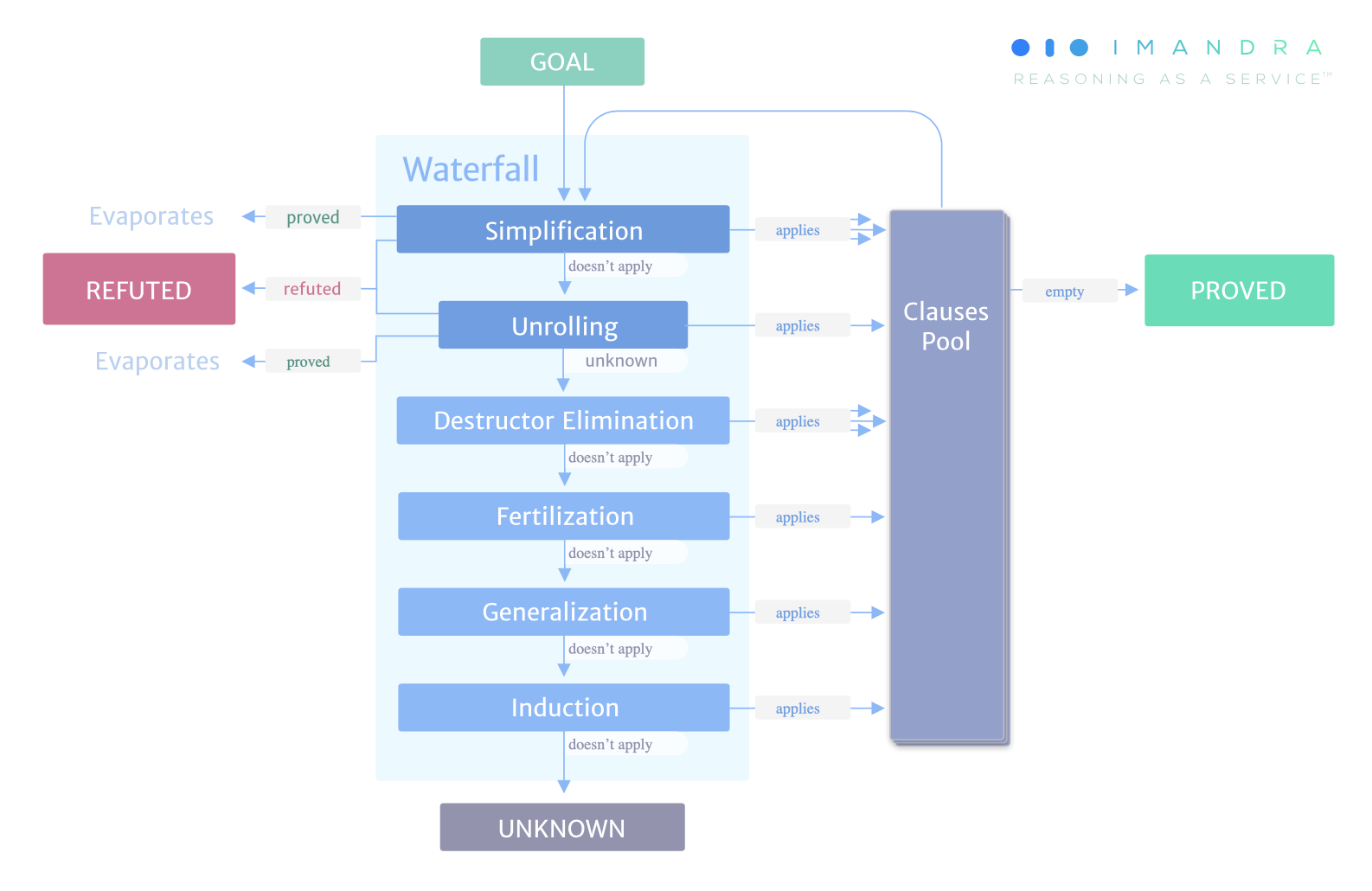}
    \caption{Imandra's inductive waterfall}
    \label{fig:screen:waterfall}
\end{figure}
\vspace{-10pt}
Imandra's waterfall contains a simplifier which automatically makes use of
previously proved lemmas.
Once proved, lemmas may be installed as rewrite, forward-chaining, elimination
or generalization rules.
Imandra gives users feedback in order to help them design efficient collections
of rules. 
With a good collection of rules (especially \lstinline{rewrite} rules), it is
hoped that ``most'' useful theorems over a given domain will be provable by
simplification alone, and induction will only be applied as a last resort.
In these cases, the subsequent waterfall moves are designed
to prepare the simplified conjecture for induction (via, e.g., generalization)
before goal-specific induction principles are synthesized.

Imandra's inductive waterfall plays an important role in what we believe
to be a robust verification strategy for applying Imandra to real-world systems.
Recall that all Imandra goals may be subjected to bounded verification
via unrolling (cf. Sec~\ref{sec:recfun}).
In practice, we almost always attack a goal by unrolling first, attempting to
verify it up to a bound before we consider trying to prove it by induction.
Typically, for real-world systems, models and conjectures will have flaws,
and unrolling will uncover many counterexamples, confusions and mistakes.
As all models are executable and all counterexamples are reflected in Imandra's
runtime, they can be directly run through models facilitating rapid investigation.
It is typically only after iterating on models and conjectures until all
(bounded) counterexamples have been eliminated that we consider trying to prove
them by induction.
Imandra's support for counterexamples also plays another important
role: as a filter on heuristic waterfall steps such as generalization.

\section{Architecture and User Interfaces}\label{sec:implementation}

Imandra is developed in OCaml and integrates with its compiler libraries.
Arbitrary OCaml code may interact with Imandra models and counterexamples
through the use of Imandra's {\tt program mode} and reflection machinery.
Imandra integrates with Z3~\cite{z3_prover} for checking satisfiability of
various classes of ground formulas. Imandra has a client-server
architecture: \begin{inparaenum}[(i)]
\item the client parses and executes models with an integrated toplevel;
\item the server, typically in the cloud, performs all reasoning.
\end{inparaenum}
Imandra's user interfaces include:
\begin{description}
  \item[Command line] for power users, with tab-completion, hints, and
    colorful messages. This interface is similar in some ways to OCaml's
    \lstinline{utop}.

  \item[Jupyter notebooks] hosted online or via local installation through
    Docker~\cite{ipython-2007}. This presents Imandra through interactive
    notebooks in the browser.

  \item[\href{https://marketplace.visualstudio.com/items?itemName=aestheticintegration.iml-vscode}{VSCode
      plugin}] where documents are checked on the fly and errors are underlined
    in the spirit of Isabelle's Prover IDE~\cite{wenzel-prover-ide}.
\end{description}
%


%
\vspace{-10pt}
\section{Conclusion}

Imandra is an industrial-strength reasoning system combining ideas from SMT,
Boyer-Moore inductive provers, and ITPs for typed higher-order logics.
Imandra delivers an extremely high degree of automation and has novel techniques
such as reflected computable counterexamples that we now believe are
indispensible for the effective industrial application of automated reasoning.
We are encouraged by Imandra's success in mainstream
finance~\cite{passmore2017formal}, and share a deep conviction that further
advances in automation and UI --- driven in large part by meeting the demands of
industrial users --- will lead to a (near-term) future in which automated
reasoning is a widely adopted foundational technology.



\bibliographystyle{plain} 
\bibliography{paper}

\end{document}